\documentclass{article}

\usepackage[final]{neurips_2025_ml4ps}

\usepackage[utf8]{inputenc} 
\usepackage[T1]{fontenc}    
\usepackage{hyperref}       
\usepackage{url}            
\usepackage{booktabs}       
\usepackage{amsfonts}       
\usepackage{nicefrac}       
\usepackage{microtype}      
\usepackage{xcolor}         
\usepackage{amsmath}
\usepackage{amssymb}
\usepackage[pdftex]{graphicx}
\usepackage{bm}
\usepackage{wrapfig}

\title{Denoising weak lensing mass maps with diffusion model and generative adversarial network}

\author{
  Shohei D. Aoyama\\
  Department of Physics\\
  Graduate School of Science\\
  Chiba University\\
  Chiba 263-8522, Japan\\
  \texttt{shohei.aoyama@chiba-u.jp}\\
  \And
  Ken Osato\\
  Center for Frontier Science\\
  Chiba University\\
  Chiba 263-8522, Japan\\
  \texttt{ken.osato@chiba-u.jp}\\
  \And
  Masato Shirasaki\\
  National Astronomical Observatory of Japan\\
  National Institutes of Natural Science\\
  Mitaka, Tokyo 181-8588, Japan\\
  \texttt{masato.shirasaki@nao.ac.jp}
}

\begin{document}

\maketitle

\begin{abstract}
The matter distribution of the Universe can be mapped through
the weak gravitational lensing (WL) effect: small distortions of the shapes of distant galaxies,
which reflects the inhomogeneity of the cosmic density field.
The most dominant contaminant in the WL effect is the shape noise;
the signal is diluted due to the finite number of source galaxies.
In order to explore the full potential of WL measurements, sharpening the signal by removing the shape noise
from the observational data, i.e., WL denoising, is a pressing issue.
Machine learning approaches, in particular, deep generative models, have proven
effective at the WL denoising task.
We implement a denoising model based on the diffusion model (DM) and conduct systematic in-depth comparisons
with generative adversarial networks (GANs), which have been applied in previous works for WL denoising.
Utilizing the large suite of mock simulations of WL observations,
we demonstrate that DM surpasses GAN in the WL denosing task in multiple aspects:
(1) the training process is more stable,
(2) taking the average of multiple samples from DM can robustly reproduce the true signal,
and (3) DM can recover various statistics with higher accuracy.

\end{abstract}

\section{Introduction}
\label{sec:introduction}
In modern cosmology, shedding light on the mysterious dark components, dark matter and dark energy,
and understanding the nature of formation and evolution of the Universe
is the most critical science goal.
Among various observations, weak gravitational lensing (WL),
which refers to small distortions of distant galaxies,
has been playing a central role to probe into the matter distribution of the Universe.
Since the galaxy shape distortion is sensitive to the gravitational potential of the foreground matter,
one can reconstruct the projected density field \citep{Kaiser1993,Schneider1995,Bartelmann1995}
through the analysis of the shapes of millions of galaxies,
and the reconstructed two-dimensional projected density field is called \textit{mass map}.
The mass reconstruction analysis is a unique and powerful approach but suffers from the shape noise;
due to the finite number of galaxies, the intrinsic shape of galaxies dilutes the WL signal.
In order to enhance the signal-to-noise ratio,
removing the shape noise from the reconstructed mass map, i.e., \textit{denoising},
is crucial for precise WL measurement.

Recently, machine learning has emerged as a promising approach as a denoising technique.
In particular, generative models in computer vision,
which take two-dimensional images as inputs and outputs,
are suitable for denoising mass maps.
For example, the generative adversarial network \citep[GAN;][]{Goodfellow2014} trained with mock WL simulations
demonstrate high performance in denoising mass maps \citep{Shirasaki2019,Whitney2024}.
In this work, we focus on the diffusion model \citep[DM;][]{Sohl-Dickstein2015},
which is one of the most powerful generative models.
DM is also classified as a deep generative model and
consists of two processes: the forward process, which adds noise to the input field,
and the reverse process, which removes the noise from the noisy data.
DM can handle various tasks relevant to image processing, such as inpainting, super-resolution, and colourisation,
and demonstrates higher performance compared with existing generative models \citep{Dhariwal2021}.

\section{Denoising mass maps with machine learning}
Our objective is to remove the shape noise from the observed mass map.
In other words, the goal is to find the optimal mapping of
the noisy mass map to the noiseless mass map.
To this end, we apply two conditional generative models: GAN and DM.
The denoising process can be regarded as the image-to-image translation problem,
where the input image is noisy observed mass maps and the output image is noiseless true mass maps.
In addition to GAN, several previous works employed DM \citep{Remy2023, Boruah2025} for denoising mass maps
motivated by the fact that DM outperforms GAN in various tasks.
Yet, a systematic comparison between GAN and DM for denoising in the same setting
has not been thoroughly addressed.
This work presents a detailed assessment of denoising performance with GAN and DM.

\textbf{Denoising with GAN} \quad
Here, we present the denoising model with \texttt{pix2pix} \citep{Isola2017} based on conditional GAN
for image-to-image translation.
The model consists of two competing networks: the generator $G$,
which produces fake denoised mass maps conditioned on noisy mass maps,
and the discriminator $D$, which distinguishes fake and true images.
For the network architecture, the generator employs U-Net \citep{Ronneberger2015},
and the discriminator utilizes four convolutional blocks
as proposed in \texttt{pix2pix}.
At the same time, the discriminator compares the input image with the target image
to determine whether the target image is a real or fake image generated by the generator.
While GANs can generate high-quality images quickly,
they suffer from various issues due to the model structure.
In general, the training of GANs is unstable due to the mode collapse and the vanishing gradient problem
compared with other existing generative models such as variational auto-encoders \citep{Kingma2014b}.
To circumvent the unstable training, alternative loss functions have been proposed,
e.g., LSGAN \citep{Mao2016} and WGAN-gp \citep{Gulrajani2017}.
However, we find that the model trained with the original loss function
of \texttt{pix2pix} yields the best results though the training process is less stable.
Therefore, we employ the original loss function as the fiducial choice.
In addition to the instability problem, the generated images by GANs are less diverse because the generator tends to ignore the latent noise term
and generate images only from the information of the conditioned images as training proceeds.
Due to this problem, \texttt{pix2pix} does not provide the latent noise term in the input data vectors.
Instead, the stochasticity appears only in dropouts of the generator network.
However, the output images are less diverse even with the treatment than other generative models.
Thus, GANs tend to fail to learn the probability distribution of the target images.

\textbf{Denoising with DM} \quad
DM \citep{Sohl-Dickstein2015,Ho2020} has been proposed
in order to circumvent the problems of GANs and enhance the quality of generated data.
It is known that the training of DM is more stable and can generate more diverse outputs
from a single input data at the cost of high computational cost.
The basic idea of DM as a generative model is to learn the process to
remove the noisy data with deep neural networks.
DM consists of two processes, the forward and reverse processes.
The forward process adds a Gaussian noise to the data
and is repeated $T$ times with the initial data.
Next, the reverse process inverts the forward process, i.e., removing noise from the noisy data.
Target data can be generated by applying the reverse process iteratively to the initial Gaussian noise.
Though the forward process is analytically tractable,
the reverse process is approximated with the deep neural networks conditioned on the input image.
In practice, for better convergence of the training process and sampling efficiency,
we adopt the simplified scheme proposed by \citet{Ho2020};
the objective function for optimisation is the reweighted variational lower-bound.
In this work, we employ \texttt{Palette} \citep{Saharia2021} implementation for image-to-image translation with DM.
The U-Net architecture is employed to model the reverse process,
and the network consists of three downsampling and upsampling layers.
For hyperparameter setting,
we find that the noise schedule has the most significant impact on denoising performance.
After testing several scheduling schemes, we employ the quadratic scheduling as the fiducial setup,
which yields the best performance.
That is presumably because the diffusion amplitude stays small during many steps and thus
the small-scale feature can be well captured by the diffusion steps.
We adopt $T = 4000$ diffusion steps in the training phase and $T = 2000$ steps in the testing phase.

\section{Results}
\textbf{Simulation data sets and training} \quad
In order to evaluate the performance of denoising with GAN and DM,
we generate realistic mock convergence maps as pairs of maps with and without shape noise.
To this end, we employ the $\kappa$TNG mock WL data suite \citep{Osato2021},
which consists of 10,000 pseudo-independent mock weak lensing maps,
where the high-resolution cosmological hydrodynamical simulation \texttt{IllustrisTNG}
\citep{Nelson2019} data are post-processed
with the multiple plane ray-tracing approach \citep{Hilbert2009}
to simulate the ray propagation in the Universe.
The mass map of $\kappa$TNG is square shaped with the area of $5 \times 5\, \mathrm{deg}^2$
and pixellated with regular $1024$ grids per dimension.
For efficiency and stability of numerical computations, we crop each map into 4 maps with equal area,
and then, we reduce the number of grids from $512$ to $256$
by average pooling with the pooling size $2 \times 2$.
To suppress transients due to instrumental systematics,
the isotropic two-dimensional Gaussian filter is applied to the maps,
and the full width at half maximum (FWHM) smoothing scale is
$\theta_\mathrm{FWHM} = 2.5\, \mathrm{arcmin}$,
which is an optimal scale for the detection of
massive galaxy clusters \citep{Miyazaki2018}.
The smoothing scale is well larger than the pixel size, i.e.,
$0.59 \, \mathrm{arcmin}$.
The final dataset consists of 40,000 smoothed maps with $2.5 \times 2.5\, \mathrm{deg}^2$
with $256$ grids per dimension.
Finally, we randomly split the entire data set into 39,000 maps for training and 1,000 maps for testing.

For training of GAN, the batch size is 1, which is shown to attain better performance
for the generator with a U-net architecture \citep[see, e.g., Appendix~6.2 in][]{Isola2017}.
The initial learning rate is set to $0.0002$, and after 100 epochs,
the learning rate decays linearly to $0$ for additional 100 epochs.
In total, the model is trained for 200 epochs.
We train the model with a single NVIDIA A100 GPU, and it took approximately 28 hours.
In contrast, testing phase, i.e., generating 1,000 denoised maps,
finishes within a few minutes.
For training of DM, the batch size is 4, and the model is trained for 85 epochs.
Similarly to the case of GAN, a single NVIDIA A100 GPU is used,
and it takes approximately 45 hours for training.
Unlike GANs, the testing for 1,000 maps takes approximately 6 hours,
i.e., 22 seconds per map with the same GPU.

Once trained with the data set, the DM denoising model can generate multiple denoised maps
from a single noisy map, which reflects the learned probability distribution function.
However, the GAN denoising model basically outputs one denoised map from one noisy map.
To virtually introduce the diversity in the GAN denoising model, we repeat training of GAN networks
five times with different initializations of network weights.
Finally, we obtain five denoised maps, each of which is generated by five trained networks for the GAN denoising model.
Note that the fluctuation among the five samples of GANs should be interpreted as the uncertainty
due to failing to find the optimal weights.
In contrast, the diversity in DM samples reflects the learned probability distribution function.

\begin{wraptable}{o}{8.5cm}
    \centering
    \vspace{-5mm}
    \caption{RMSE and Pearson correlation coefficients $\rho$ for the denoised maps with GAN and DM.
    These quantities are measured for five individual samples
    and their mean and median.
    The metrics computed from noisy maps before applying denoising are also shown as ``No denoising''.}
    \vspace{2mm}
    \begin{tabular}{l|cc|cc}
        \hline
        & \multicolumn{2}{c|}{RMSE ($\times 10^{-2}$) $\downarrow$} & \multicolumn{2}{c}{Pearson coef. $\rho$ $\uparrow$} \\
        5 samples & GAN & DM & GAN & DM \\
        \hline
        \hline
        Individual & 1.12 & 1.11 & 0.644 & 0.638\\
        Mean & 0.87 & \textbf{0.86} & \textbf{0.758} & 0.757 \\
        Median & 0.90 & 0.89 & 0.743 & 0.742 \\
        \hline
        \hline
        No denoising & \multicolumn{2}{c|}{1.47} & \multicolumn{2}{c}{0.67}\\
        \hline
    \end{tabular}
    \label{tab:RMSE_Pearson}
\end{wraptable}

\begin{figure}[!htbp]
  \centering
  \includegraphics[width=0.9\linewidth]{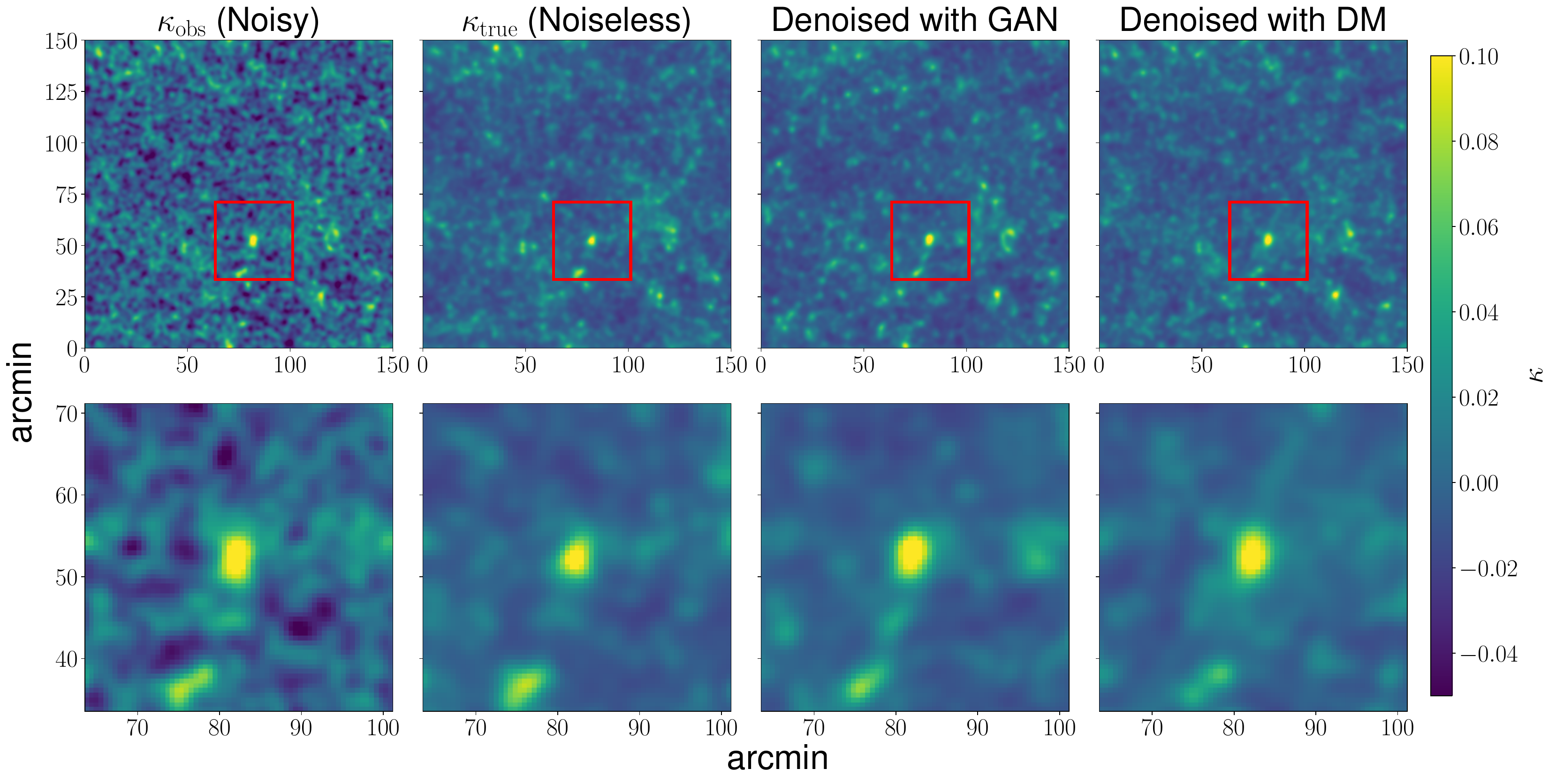}
  \caption{\textit{Upper row}: Comparison between the denoised convergence maps
  with GAN and DM and the ground truth map.
  \textit{Lower row}: Zoom-in maps around the peak with the highest significance in the true noiseless map.
  The red squares in the maps in the upper row indicate
  the corresponding zoom-in regions displayed in the lower row.}
  \label{fig:maps}
\end{figure}

\textbf{Pixel-level comparison} \quad
We apply the GAN and DM denoising models to 1,000 test maps.
For each test map, we sample five denoised maps from five networks for GAN or
by sampling with five different initial Gaussian noise for DM. 
Figure~\ref{fig:maps} illustrates one example of denoising results:
the input noisy map, the ground truth map, and the denoised maps with GAN and DM.
The small-scale noisy feature is successfully removed in the denoised maps.
We measure the root mean square error (RMSE) and the Pearson correlation coefficient $\rho$
at the pixel level for quantitative comparisons.
Table~\ref{tab:RMSE_Pearson} shows the RMSE and Pearson coefficients
of 1,000 mean or median maps of five denoising realisations
and 5,000 individual maps estimated by GAN and DM.
For comparison, the metrics without denoising are also shown.
RMSE significantly improves by denoising with GAN or DM,
which demonstrates the effectiveness of the denoising methods.
On the other hand, the Pearson coefficient computed from individual maps stays similar even after denoising.
Taking mean or median of the maps among the five realisations results in improvement for both metrics.
In terms of comparison between GAN and DM, the difference is considerably small.

\textbf{Statistics} \quad
Next, we investigate the cosmological statistics to quantify the performance of denoising.
We begin with the angular power spectrum $C (\ell)$, which is defined as
\begin{equation}
C(\ell) = \sum_{\ell - \Delta \ell/2< \ell' < \ell + \Delta \ell/2} |\tilde{\kappa}(\bm{\ell}')|^2
/ \sum_{\ell - \Delta \ell/2 < \ell' < \ell + \Delta \ell/2} 1,
\end{equation}
where $\bm{\ell}$ and $\tilde{\kappa} (\bm{\ell})$ denote the wave-vector and the mass map in Fourier space, respectively,
and $\Delta \ell$ is the width of binning. We set the wave-vector bin range as $[100, 10000]$
with logarithmically equally spaced bins.
The power spectrum reflects the fluctuation amplitude at the angular scale of $\sim \pi / \ell$.
We also measure the one-point probability density function (PDF) of the denoised mass maps,
which also contain cosmological information and can be easily measured from data \citep{Thiele2023}.

\begin{figure}
  \centering
  \includegraphics[width=0.48\linewidth]{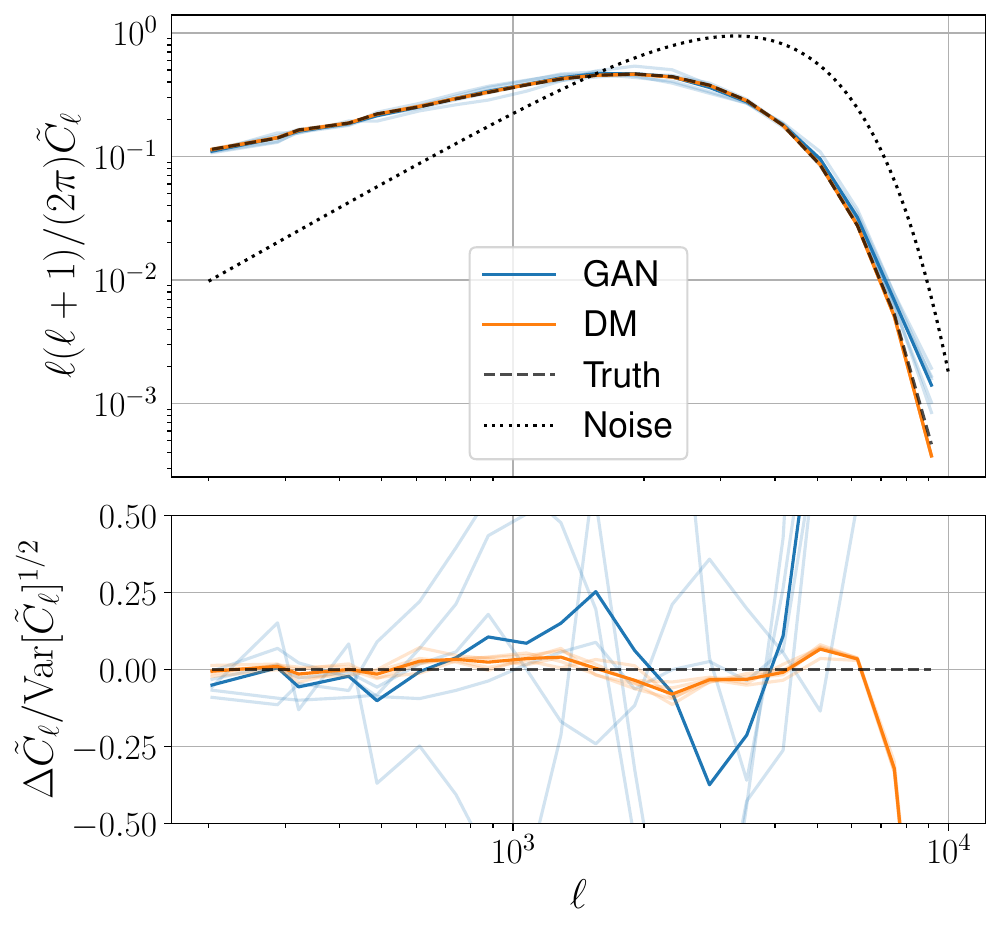}
  \includegraphics[width=0.48\linewidth]{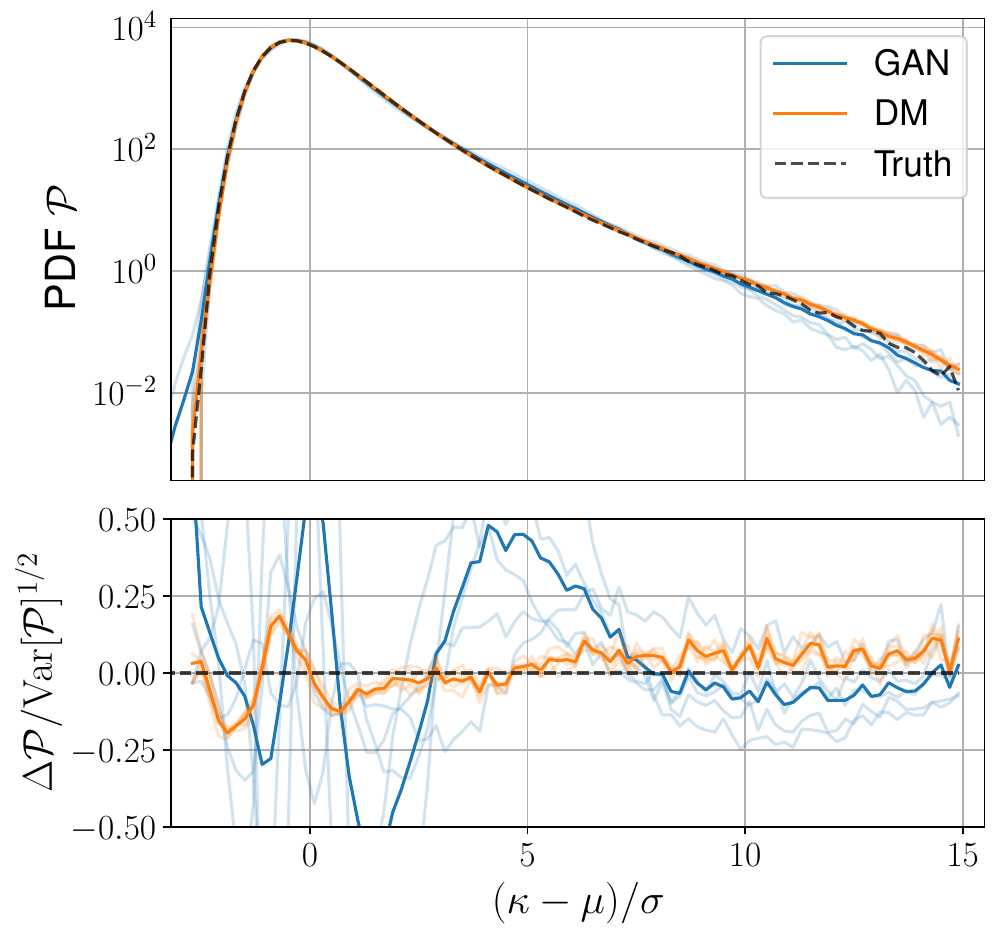}
  \caption{\textit{Left panel}: The angular power spectra of the normalised ground-truth maps (dashed line)
  and denoised maps (solid lines) with GAN and DM.
  The five blue (orange) thin lines correspond to the five networks (samples) from GAN (DM).
  The thick blue (orange) line shows the average power spectrum for GAN (DM).
  The dotted line corresponds to the noise power spectrum.
  The lower panel shows the fractional difference.
  \textit{Right panel}: PDFs of the denoised maps with GAN (blue) and DM (orange).
  The maps are normalised so that the PDF has zero mean and unit variance.}
  \label{fig:stats}
\end{figure}

Figure~\ref{fig:stats} shows the power spectra of noiseless convergence maps and denoised maps with GAN and DM,
and the power spectrum of each sample is shown as thin lines.
The lower panel illustrates the fractional difference normalized by the standard deviation of statistics of true maps.
Overall, DM outperforms GAN with regard to power spectrum reconstruction;
DM can reproduce the power spectrum within $0.1$ in terms of the fractional difference for $\ell \lesssim 6000$.
Though the noise power spectrum dominates for $\ell \gtrsim 2000$,
DM accurately reconstructs the power spectrum down to small scales,
which would be inaccessible without denoising.
On the other hand, the range where GAN reproduces the correct power spectrum
is limited to the large scale ($\ell \lesssim 1000$).
Furthermore, the variance among the five networks for GAN is quite large even at the intermediate scale ($\ell \sim 1000$).
Averaging over five samples results in a reasonable estimate of the power spectrum.
In contrast, the five samples of DM indicate less variance;
all five samples perform well up to $\ell \simeq 6000$.

A similar trend is found in PDF.
The reconstruction accuracy of DM is at least $0.1$ in terms of the fractional difference for the entire range.
All five samples show a similar trend, and
this consistency demonstrates the robustness of the sampling process inherent to the DM.
On the other hand, the results of GAN yield larger variability among the five trained networks.
We have measured other statistics such as angular bispectrum and scattering transform \citep{Cheng2020}
and found that DM can outperform GAN in all the statistics we addressed.

\section{Conclusions}
In order to enhance the signal-to-noise ratio of WL measurement,
denoising the shape noise based on machine learning is an effective approach.
In particular, deep generative models such as GAN and DM are ideal methods for such denoising tasks
because these methods have been widely applied in image processing.
In this work, we explore the capabilities of GAN and DM in the specific task of denoising WL mass maps
and conduct systematic fair comparisons utilizing the large suite of mock WL simulations.
We reveal that the performance of DM is superior to GAN both at the pixel level and in statistics.
Furthermore, the large variation of GAN samples results in less accurate statistics.
In contrast, statistics from the samples taken from the DM denoising model
are more robust and taking the mean or mode of five samples
yields the unbiased reconstruction of mass maps.

\bibliographystyle{plainnat}
\bibliography{main}

\end{document}